\documentclass[trackchanges]{aastex701}
\usepackage{amsmath}

\begin{document}

\title{2MASS Re-processing I: The Search for Faint Objects}

\author{Zehao Liu}
\affiliation{Shanghai Astronomical Observatory, Chinese Academy of Sciences, Shanghai 200030, China}
\affiliation{University of Chinese Academy of Sciences, Beijing 100049, China}
\email{liuzehao@shao.ac.cn}

\author{Xiyan Peng}
\affiliation{Shanghai Astronomical Observatory, Chinese Academy of Sciences, Shanghai 200030, China}
\email{xypeng@shao.ac.cn}

\correspondingauthor{Xiyan Peng}
\email{xypeng@shao.ac.cn}

\author{Zhenghong Tang}
\affiliation{Shanghai Astronomical Observatory, Chinese Academy of Sciences, Shanghai 200030, China}
\affiliation{University of Chinese Academy of Sciences, Beijing 100049, China}
\email{zhtang@shao.ac.cn}

\author{Zhaoxiang Qi}
\affiliation{Shanghai Astronomical Observatory, Chinese Academy of Sciences, Shanghai 200030, China}
\affiliation{University of Chinese Academy of Sciences, Beijing 100049, China}
\email{zxqi@shao.ac.cn}

\author{Shilong Liao}
\affiliation{Shanghai Astronomical Observatory, Chinese Academy of Sciences, Shanghai 200030, China}
\affiliation{University of Chinese Academy of Sciences, Beijing 100049, China}
\email{shilongliao@shao.ac.cn}

\author{Yong Yu}
\affiliation{Shanghai Astronomical Observatory, Chinese Academy of Sciences, Shanghai 200030, China}
\affiliation{University of Chinese Academy of Sciences, Beijing 100049, China}
\email{yuy@shao.ac.cn}


\begin{abstract}
We present an automated, DAOFind-based pipeline developed to reprocess J-band Atlas All Sky Release Survey Images from the Two Micron All Sky Survey (2MASS). By optimizing the detection parameters and implementing a screening procedure that jointly evaluates the signal-to-noise ratio and central sharpness, the pipeline effectively identifies faint point sources that were previously undetected. Applying this method to eight representative sky regions improves the 2MASS detection limit from 16.20 to 16.60 mag and increases the number of detected point sources by approximately $21.36\%$ relative to the 2MASS Point Source Catalog, with a false-positive rate of only $4.80\%$. These results demonstrate that the proposed reprocessing pipeline can substantially enhance the scientific yield of archival 2MASS data, providing valuable faint-source supplements for studies of time-domain variability, Galactic structure, and cold, low-luminosity objects.
\end{abstract}



\section{INTRODUCTION} 
2MASS \citep{skrutskie2006two} was the first ground-based survey to map the entire sky in the near-infrared. It employed two dedicated 1.3 m telescopes located at Mount Hopkins, Arizona, and Cerro Tololo, Chile, and operated between June 1997 and February 2001. The survey covered almost the entire celestial sphere in three photometric bands—J (1.25 $\mu$m), H (1.65 $\mu$m), and Ks (2.16 $\mu$m)—and produced two major catalogs: The project produced a Point Source Catalog containing 470,992,970 objects \citep{cutri20032mass}, an Extended Source Catalog with 1,647,599 entries \citep{jarrett20002mass}, and 4,121,439 FITS images covering $99.998\%$ of the sky, establishing an indispensable near-infrared reference for astronomical research \citep{skrutskie2006two}. To ensure the reliability of its products, the 2MASS team adopted a deliberately conservative data-processing strategy. The 2MASS Point Source Catalog is essentially limited at signal-to-noise ratios (SNR) $\approx 10$, with only a small fraction of sources fainter than this threshold being retained. This cautious approach inevitably led to the omission of many faint sources near the detection limit. To fully exploit the scientific information contained in the original 2MASS observational images and provide valuable survey data for long-term time-domain astronomy, we have improved the source detection algorithm and reprocessed the 2MASS images. This approach aims to maximize the detection of faint point sources captured in the images. These previously uncataloged “new confirmed point sources" constitute an important supplement to the 2MASS point source catalog.

The roughly two-decade temporal baseline between 2MASS and more recent surveys—such as the UKIRT Infrared Deep Sky Survey \citep{lawrence2007ukirt}, the VISTA Variables in the Via Lactea \citep{minniti2010vista,saito2012vvv}, the Wide-field Infrared Survey Explorer and its NEOWISE reactivation mission \citep{wright2010wide,mainzer2014initial}, the DESI Legacy Imaging Surveys ~\citep{dey2019overview}, the Euclid mission \citep{mellier2024euclid} and the Gaia mission \citep{vallenari2023gaia}—provides a unique opportunity to exploit the faint sources recovered from 2MASS data for a wide range of scientific applications. First, the long time baseline enables the construction of extended light curves that are invaluable for studying the long-term evolution of variable stars, eruptive young stellar objects, and active galactic nuclei \citep{carpenter2001near,plavchan2008near,green2024increasing,zaharieva2025identification}. Second, the near-infrared sensitivity of 2MASS facilitates the detection of cool and heavily reddened sources. Systematic recovery of these faint objects mitigates the incompleteness of optical surveys for low-temperature dwarfs (e.g., L/T types) and nearby low-mass stars, thereby improving the completeness of stellar census in the solar neighborhood \citep{kirkpatrick1999dwarfs,burgasser1999discovery,kirkpatrick200067,kirkpatrick2021field,dal2023ultracool}. In addition, cross-matching the reprocessed 2MASS data with modern high-precision astrometric catalogs enables more accurate and complete proper-motion determinations \citep{roeser2010ppmxl,theissen2016motion,tian2017gaia,tian2020extended,kirkpatrick2014allwise,kirkpatrick2016allwise,marrese2019gaia}. Combining proper-motion, parallax, and photometric information further allows statistical separation of the thin-disk, thick-disk, and halo populations of the Milky Way, providing new insights into Galactic structure and kinematics \citep{smith2018virac,ahmed2024deep}).

This paper focuses on methods for detecting faint sources in 2MASS J-band Atlas All Sky Release Survey Images \citep{cutri2003explanatory}, hereafter referred to as “Atlas Images". A principal challenge in this work is the reliable discrimination between pixels containing astrophysical sources and those dominated by background emission or noise \citep{masias2012review}. To meet this challenge, we have developed a specialized automated pipeline for faint sources detection, designed with the following objectives: 1) to prioritize the detection of point sources; 2) to identify as many \textbf{new confirmed faint sources} as possible; and 3) to maintain an acceptably low false-positive rate across diverse sky regions.

The pipeline processes the Atlas images, which constitute the primary science-ready data products from 2MASS. Each image results from the co-addition of multiple short exposures obtained in a microscanning pattern and has undergone rigorous flat-fielding, sky subtraction, and astrometric and photometric calibration \citep{skrutskie2006two,cutri2003explanatory}. Additionally, overlapping areas exist between adjacent Atlas images. With a pixel scale of $1^{\prime\prime}$, these images provide a stable, well calibrated foundation. We directly utilize key header parameters \citep{cutri2003explanatory}—specifically the background noise estimate ($SKYSIG$) and the seeing shape parameter ($SEESH$)—to optimize the detection of faint sources.

To validate the authenticity of the pipeline detections, we performed external checks using the 2MASS Point Source Catalog (PSC) and the VISTA Hemisphere Survey (VHS) band-merged multi-waveband catalog. Although VHS is confined to the southern hemisphere, its passbands match those of 2MASS and reach depths about 30 times greater in $J$ and $K_s$, with superior spatial resolution. The VHS $J$-band survey covers 16,689 deg$^2$ \citep{Pons2020VHS_DR5}, enabling direct positional cross-validation of new detected faint sources in the same bands. Consequently, all Atlas images used here are drawn from overlapping southern-sky regions common to both 2MASS and VHS.

In fact, the 2MASS catalogs related to point sources are not limited to the PSC. There also exist two additional types of point source catalogs \citep{cutri2003explanatory}: the 2MASS Survey Point Source Reject Table (PSRT, \citealt{PSRT}) and the 2MASS 6X w/LMC/SMC Point Source Working Database/Catalog (6x-PSWDB/6x-PSC, \citealt{cutri2012vizier}). The PSRT contains lower SNR and less reliable sources from the 59,731 scans used to construct the PSC, as well as all sources from the 10,981 scans not selected for the full-sky release. These include reliable detections of real astrophysical sources, spurious extractions of low SNR events, image artifacts and transients such as cosmic rays and meteor trails. Furthermore, the 6x-PSC is a point source catalog obtained through a special observing strategy implemented during the final year of the 2MASS observatory operations. This strategy provided exposure times six times longer than those of the main 2MASS survey measurements. The 2MASS “6x" observations achieved sensitivities $\sim1$ mag deeper than the main 2MASS survey, and covered approximately 590 $deg^2$ of sky in 30 discrete regions. Given that the observational images used for the 6x-PSC are not consistent with the Atlas images employed in this study, in the section 3 of this paper we therefore compare only the PSRT data with our pipeline data to analyze the differences between the two source datasets in the low SNR regime.

The remainder of this paper is structured as follows. Section 2 elaborates on the faint sources detection pipeline and defines the detection criteria for key procedural steps. Section 3 analyzes the detection results from the eight sample sky regions and compares these results with the PSRT data; Finally, Section 4 presents the conclusions and future prospects.

\section{DETECTION PIPELINE} 
This section systematically outlines the pipeline for detecting faint sources in Atlas images and elaborates on the methodology used to determine its key parameters. To ensure the stability and robustness of the pipeline when processing images with varying backgrounds and stellar densities, we selected eight distinct sky regions (see Table \ref{tab:1} for their properties) and randomly sampled one Atlas image from each region for comprehensive pipeline optimization and testing.

\begin{deluxetable*}{rllp{5cm}}
\tablewidth{0pt}
\tablecaption{Eight sky regions information \label{tab:1}}
\tablehead{
\colhead{ID} & \colhead{Frame Name} & \colhead{Bounds (RA.,DEC.)} & \colhead{Region Description}
}
\startdata
\hline
1 & asky\_980916s1020056 & $[10.5^\circ,11.5^\circ]$, $[-2.0^\circ,-1.0^\circ]$ & High Galactic latitude regions exhibit lower stellar densities and a higher prevalence of background galaxies. \\
\hline
2 & asky\_981023s1060092 & $[60.0^\circ,62.9^\circ]$, $[-70.1^\circ,-69.1^\circ]$ & Intermediate Galactic latitude regions, located near the Large Magellanic Cloud. \\
\hline
3 & asky\_000422s1160267 & $[0.0^\circ,50.0^\circ]$$\cup$$[270.0^\circ,360.0^\circ]$, $[-89.8^\circ,-89.0^\circ]$ & Located near the South Celestial Pole. \\
\hline
4 & asky\_001124n0860056 & $[83.7^\circ,84.7^\circ]$, $[-5.4^\circ,-4.4^\circ]$ & Located above the Orion Nebula region, this area contains abundant bright interstellar medium. \\
\hline
5 & asky\_980514s0120221 & $[200.8^\circ,202.2^\circ]$, $[-43.0^\circ,-42.0^\circ]$ & Situated in the peripheral region of NGC 5128. \\
\hline
6 & asky\_990416s0560198 & $[210.0^\circ,211.3^\circ]$, $[-40.4^\circ,-39.4^\circ]$ & A low Galactic latitude region. \\
\hline
7 & asky\_000529s1240009 & $[287.3^\circ,289.3^\circ]$, $[-60.6^\circ,-59.6^\circ]$ & Located within NGC 6752, this region is characterized by a high stellar density. \\
\hline
8 & asky\_000310s0540127 & $[194.2^\circ,197.0^\circ]$, $[-69.7^\circ,-68.7^\circ]$ & This region lies in the vicinity of the Galactic plane and is characterized by a high stellar density. \\
\hline
\enddata
\end{deluxetable*}

\subsection{Overview of the detection pipeline}
The pipeline includes procedures such as the preprocessing of single Atlas images, the detection and screening of sources, and the deduplication of repeated detections in overlapping fields.

First, before detection, the header information and data of the Atlas images are read, and background subtraction is performed using the Background2D algorithm from the Photutils \citep{larry_bradley_2025_14889440}. Subsequently, the DAOStarFinder is employed for preliminary target detection on the images, with the aim of detecting as many faint targets as possible in the Atlas images. DAOStarFinder (abbreviated as “DAOFind") originates from the preliminary star-finding routine in DAOPHOT \citep{stetson1987daophot}, developed and maintained by the Dominion Astronomical Observatory in Canada. To eliminate incomplete stellar images at the edges of the images, detected targets within 26 pixels from the top and bottom edges and within 36 pixels from the left and right edges of the Atlas images are excluded from subsequent processing. Since the overlapping regions between adjacent Atlas images are larger than the above ranges, this operation will not cause any omission in the survey area.

Although only detections with a SNR above the predefined threshold were retained, a significant number of false positives—accounting for approximately 5\% to 40\% of the total sources—were still present in the detection results across different Atlas images. To mitigate this high false positive rate, secondary screening strategies, tailored to the stellar density of each field, were applied to filter the faint sources.

Finally, for duplicate sources detected in overlapping regions of the Atlas images, the KD-tree method \citep{2020SciPy-NMeth} is employed to identify and remove repeated entries, ensuring a clean list of unique sources.

\subsection{Parameter optimization}

The determination of parameters for key steps in the pipeline requires cross-verification using external star catalogs: by cross-matching the detection results from each step with the catalogs, key metrics such as the total number of detections, the number of new confirmed point sources, or the count of false positives are statistically analyzed and compared, thereby identifying the optimal parameter combination.

Catalog matching refers to the process of cross-referencing astronomical catalogs from different observational sources—such as telescopes, wavebands, time-domain surveys, or sky survey programs—to determine whether recorded observations correspond to the same celestial object. In this study, we utilize the 2MASS point source catalog and the VHS catalog as external references, cross-matching them with our detected sources using matching radii of $1^{\prime\prime}$ and $2^{\prime\prime}$, respectively. This matching process is designed to address two key questions: 1) whether the detected objects are astrophysical sources, and 2) whether these confirmed sources represent new confirmed point sources relative to the PSC. Here, authentic sources are defined as detections that match either a 2MASS point source or a VHS entry. Among these, sources that lack a 2MASS counterpart but have a match in VHS are classified as new confirmed sources, while unmatched detections are categorized as false positives. For the classification of point sources and extended sources among new confirmed sources, the MERGEDCLASS parameter from the VHS catalog is used for determination \citep{Pons2020VHS_DR5}. This parameter categorizes VHS sources into four classes: galaxy, star, noise, and saturated source. The data obtained through the aforementioned matching process can then be used to calculate the point source growth rate and the false alarm rate, which are defined as shown in Equations 1 and 2.

\begin{equation}
\mathrm{GR}(\%) = \frac{N_{\mathrm{new}}}{N_{\mathrm{2MASS}}}
\end{equation}

$GR(\%)$ represents the relative increase in the number of confirmed point sources compared to the PSC. Here, $N_{\mathrm{new}}$ refers the number of new confirmed sources, while $N_{\mathrm{2MASS}}$ denotes the number of point sources from the PSC within the detection region.

\begin{equation}
\mathrm{FP}(\%) = \frac{N_{\mathrm{FP}}}{N_{\mathrm{DAOFind}}}
\end{equation}

$\mathrm{FP}(\%)$ represents the false positive rate. $N_{\mathrm{FP}}$ denotes the number of detected objects that match neither sources in the 2MASS point source catalog nor those in the VHS catalog. $N_{\mathrm{DAOFind}}$ refers to the total number of sources detected by DAOFind. Based on the statistical results obtained from matching with different parameters, we proceed to determine the optimal conditioning parameters for key steps in the detection pipeline.

\subsubsection{Determination of DAOFind detection parameters}
Among the DAOFind parameters, the most critical are the detection threshold and the detection full width at half maximum ($\mathrm{FWHM}_{\mathrm{det}}$). The detection threshold defines the minimum confidence level required to classify a detected object as valid, and it will be adjusted through the SKYSIG parameter. For the $\mathrm{FWHM}_{\mathrm{det}}$, it is determined based on the empirical relationship between the seeing and the image full width at half maximum, as given in Equation 3 \citep{Cutri1998Seeing}, the value of the image full width at half maximum ($\mathrm{FWHM}_{\mathrm{img}}$)  is determined by the $SEESH$ provided in the Atlas image file header.

\begin{equation}
\mathrm{FWHM}_{\mathrm{img}} \ \mathrm{(arcsec)} = 3.13 \times \mathrm{SEESH} - 0.46
\end{equation}

The relationship between $\mathrm{FWHM}_{\mathrm{img}}$ and $\mathrm{FWHM}_{\mathrm{det}}$ is established using Equation 4, where $\mathrm{n}_{\mathrm{1}}$ is the conversion coefficient.

\begin{equation}
\mathrm{FWHM}_{\mathrm{det}} = \mathrm{n}_{\mathrm{1}} \times \mathrm{FWHM}_{\mathrm{img}}
\end{equation}

Additionally, the relationship between the threshold and $SKYSIG$ is given by Equation 5, where $\mathrm{n}_{\mathrm{2}}$ is also a conversion coefficient.

\begin{equation}
threshold = \mathrm{n}_{\mathrm{2}} \times SKYSIG
\end{equation}

We performed source detection on the Atlas images using different conversion factors ($\mathrm{n}_{\mathrm{1}}$, $\mathrm{n}_{\mathrm{2}}$) and calculated the SNR of the detections. The SNR in this work is calculated with the following formula:

\begin{equation}
\mathrm{SNR} = \frac{\mathrm{flux}}{\mathrm{fluxerr}} = \frac{f_{\sigma} - \pi\sigma^{2} \, \mathrm{SKYSIG}}{\sqrt{\pi\sigma^{2} \times \mathrm{SKYSIG}^{2}}}
\end{equation}

\begin{figure*}[ht!]
\centering
\scalebox{0.22}{\includegraphics{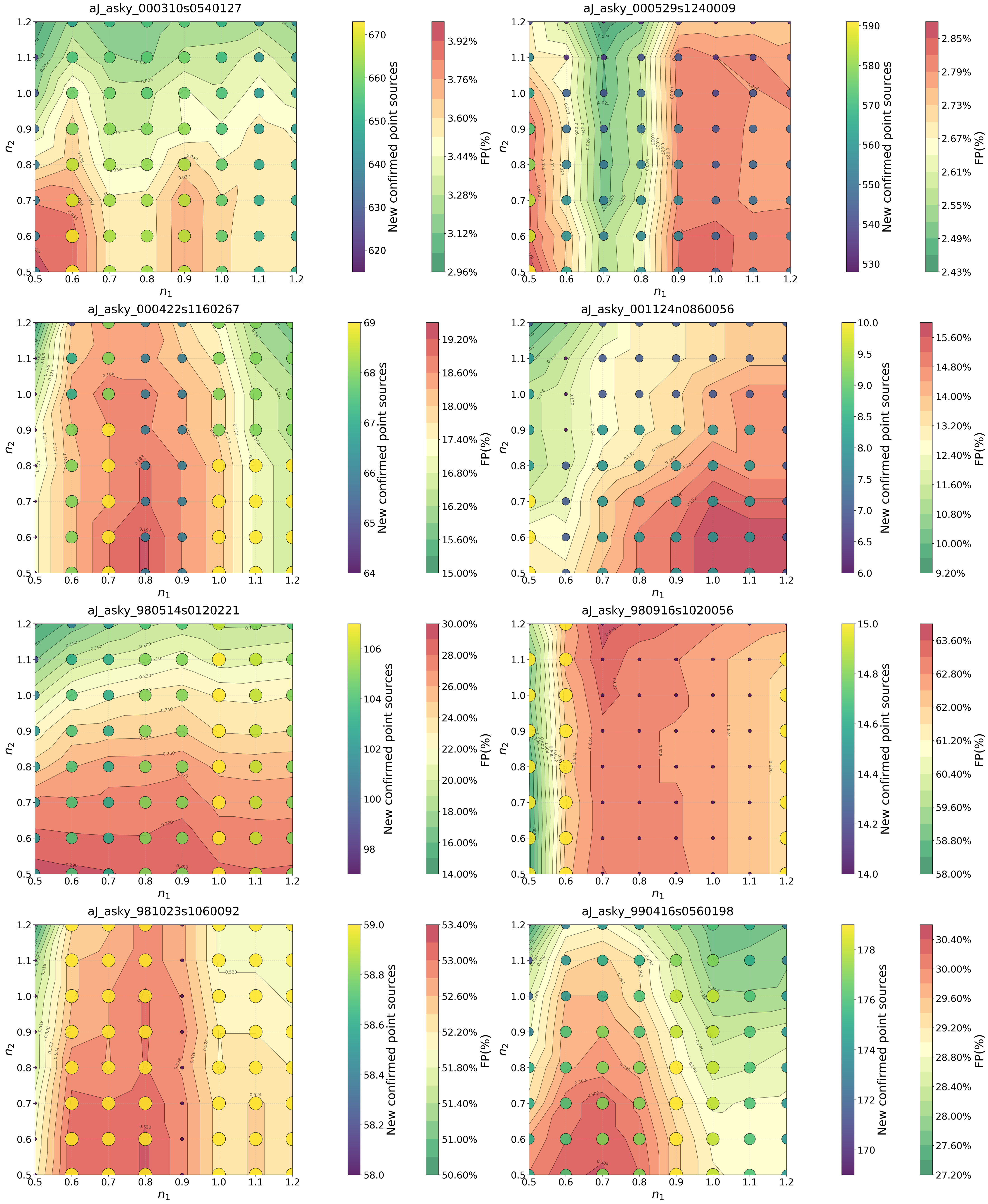}}
\caption{The figure shows the variation trends in the number of newly added point sources with a SNR greater than 3 and the false positive rate for eight Atlas images under different detection parameters. The horizontal axis represents the value of the conversion coefficient $\mathrm{n}_{\mathrm{1}}$ for $\mathrm{FWHM}_{\mathrm{det}}$, and the vertical axis represents the value of the conversion coefficient $\mathrm{n}_{\mathrm{2}}$ for the detection threshold. In the figure, variations in the number of new confirmed point sources are represented by different colors and data point sizes, while changes in the false alarm rate are indicated by color gradients and contour lines.
\label{fig:1}}
\end{figure*}

Here, $\sigma = \mathrm{FWHM}_{\mathrm{img}}/(2\sqrt{2\ln 2})$ is used as the photometric radius. This aperture photometry method effectively prevents the flux measurement of faint point sources from being contaminated by nearby bright stars, without compromising the detection of normal or bright sources. The value $\mathrm{f}_{\mathrm{\sigma}}$ represents the background-subtracted flux within this aperture. To mitigate potential flux overestimation for faint sources in cases of complex backgrounds in Atlas images, we apply a further correction to $\mathrm{f}_{\mathrm{\sigma}}$ using the filtered noise term, $\pi\sigma^{2} \times \mathrm{SKYSIG}$, estimated within the source region. The parameter fluxerr represents the noise value, calculated with reference to the method used in SExtractor \citep{bertin1996sextractor}. We then retained only those detections with a SNR greater than 3 for additional screening. The detection results are presented in Figure \ref{fig:1}.

Figure 2.1 shows the variation trends in the number of new confirmed point sources with SNR greater than 3 and the false positive rate when $\mathrm{n}_{\mathrm{1}}$ and $\mathrm{n}_{\mathrm{2}}$ take the values $[0.5, 0.6, 0.7, 0.8, 0.9, 1.0, 1.1, 1.2]$, respectively. As can be seen from the figure, in the vicinity of $\mathrm{n}_{\mathrm{1}} = 0.7$ and $\mathrm{n}_{\mathrm{2}} = 1.0$, the dense stellar fields (Regions 7 and 8) exhibit a relatively high number of new confirmed point sources and a low false positive rate. Although the results obtained for sparse stellar fields near these parameters are not entirely consistent, the difference between the number of new confirmed point sources at this point and the maximum number of new confirmed point sources in the same region is only about $1\%$, and the false positive rates mostly fall around the median, which has little impact on the overall detection results of the images. Therefore, in most sky regions, the number of new confirmed point sources with SNR greater than 3 and the false alarm rate obtained near $\mathrm{n}_{\mathrm{1}} = 0.7$ and $\mathrm{n}_{\mathrm{2}} = 1.0$ show comparative advantages. Thus, the DAOFind detection parameters can be set as: $threshold = 1.0 \times SKYSIG$, $\mathrm{FWHM}_{\mathrm{det}} = 0.7 \times \mathrm{FWHM}_{\mathrm{img}}$. Of course, the above is only a preliminary detection, aimed solely at selecting a unified set of DAOFind parameter conversion coefficients for different sky regions. Further screening of the detected targets will be conducted subsequently.

\subsubsection{Determination of the central sharpness selection criterion}

Following preliminary detection of Atlas images using DAOFind, we observed that while a large number of new confirmed point sources can be detected across different images, the false positive rate in corresponding regions remains high. To further reduce the false positive rate, inspired by the local peak search method \citep{masias2012review}, this paper introduces an approach for screening faint sources based on the central sharpness value $\Delta f$. This sharpness metric aims to evaluate how closely a faint detection source resembles an ideal point source, comprehensively reflecting the energy concentration characteristics of the source at the resolution limit. Specifically, a higher sharpness value indicates that the spatial distribution of the source more closely approximates a point source. Accordingly, a secondary screening of low SNR sources within different stellar density fields is performed to further reduce the false positive rate. The central sharpness value is given by Equation 7,

\begin{equation}
\Delta f = f_{\mathrm{center}} - \Sigma
\end{equation}

\begin{figure*}[ht!]
\centering
\scalebox{0.8}{\includegraphics{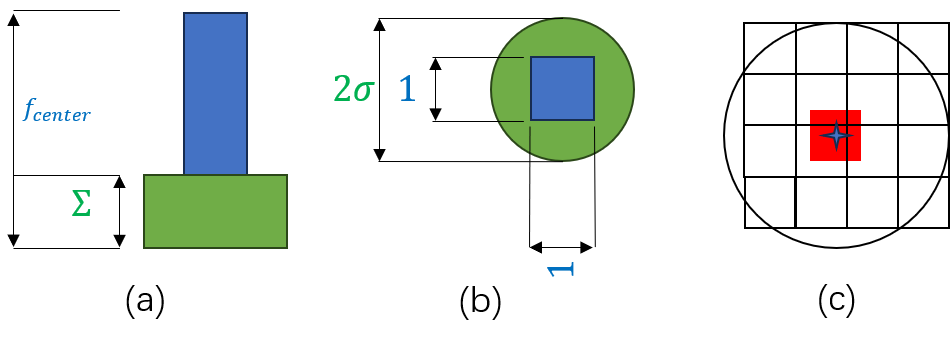}}
\caption{(a) Schematic side view of a faint source showing the peak flux and its surrounding region; (b) Top view of the faint source. Panels (a) and (b) facilitate the interpretation of Equations 7 and 8, highlighting the peak flux $f_{\mathrm{center}}$ (blue) and the flux surface density $\sigma$ (green). (c) Spatial relationship between the $1\times1$ pixel region used for peak flux calculation (red) and other pixels of the detected source.
\label{fig:2}}
\end{figure*}

Here, $f_{\mathrm{center}}$ and $\Sigma$ represent the central peak pixel flux of the detected source and the flux surface density in its vicinity (see Figure \ref{fig:2}; $\Sigma$ is calculated by Equation 8), respectively. $\Delta f$ is used to characterize the sharpness of the detected source. The definition of $\sigma$ is the same as in Equation 6.

\begin{equation}
\Sigma = \frac{f_{\sigma} - f_{\mathrm{center}}}{2\pi\sigma^{2} - 1}
\end{equation}

\begin{figure*}[ht!]
\centering
\scalebox{0.20}{\includegraphics{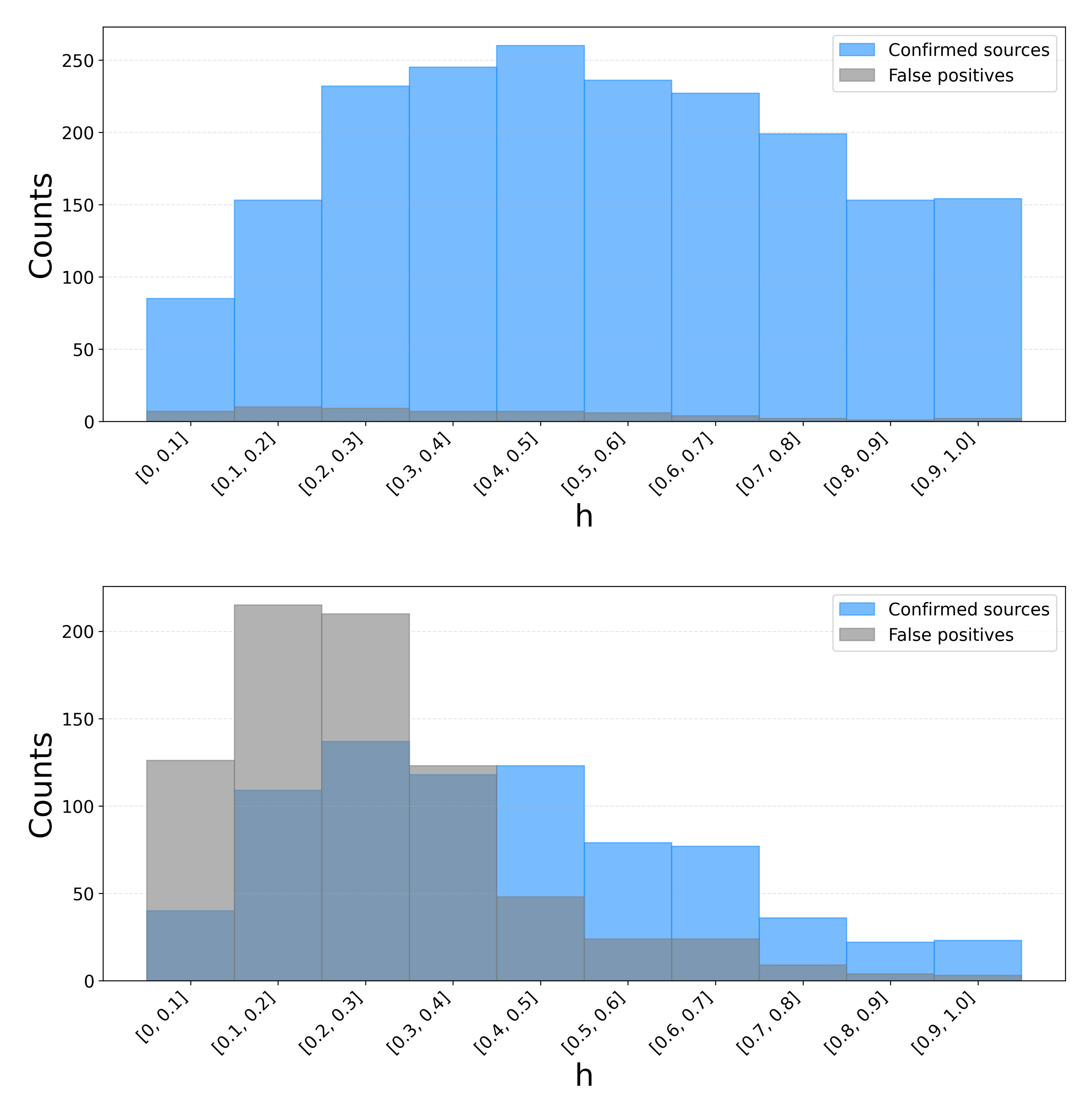}}
\caption{ $\mathrm{SNR} \leq 10$ ($h \in (0,1]$ with a step size of 0.1), while the y-axis quantifies detected sources by type.
\label{fig:3}}
\end{figure*}

\begin{figure*}[ht!]
\plotone{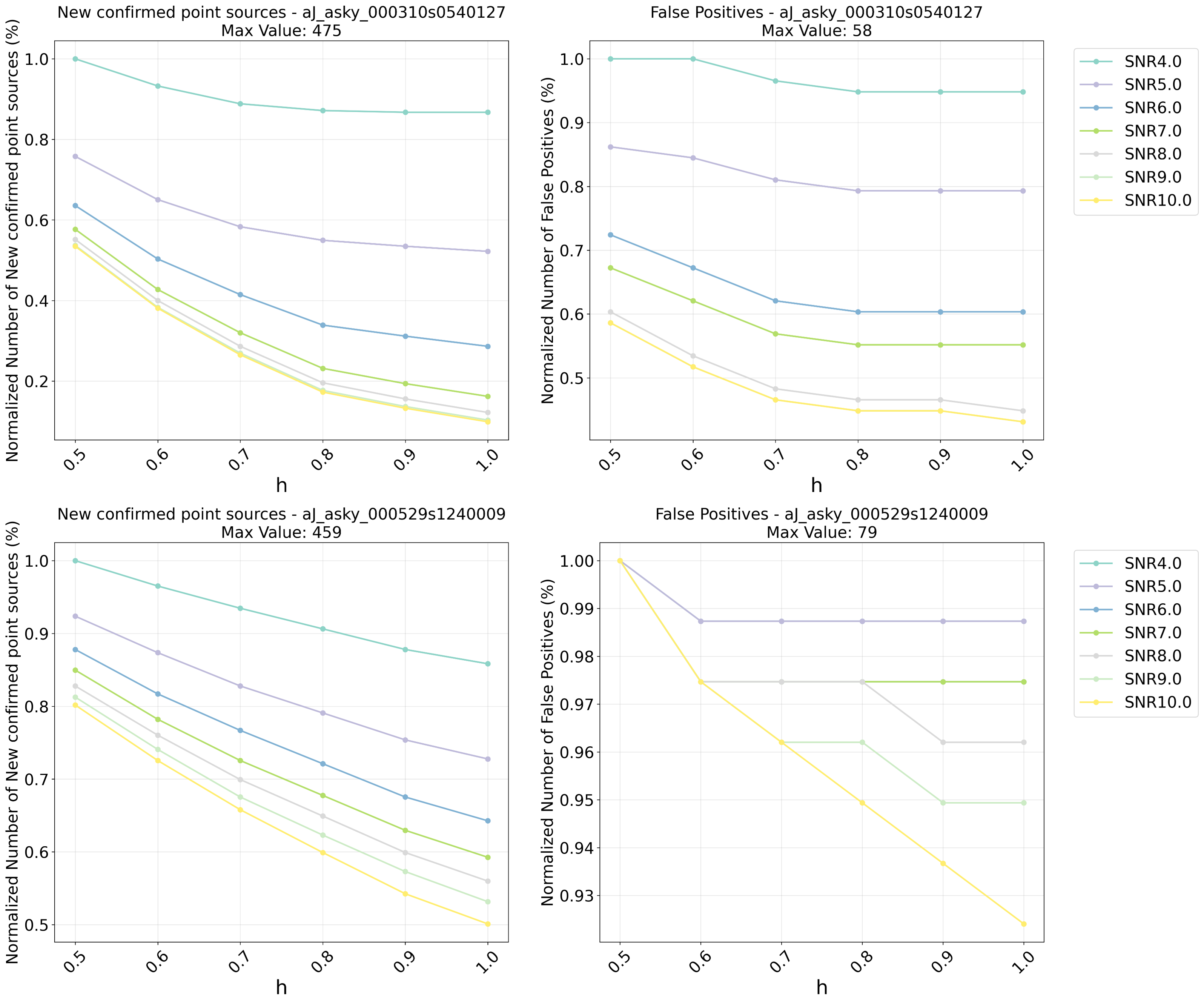}
\caption{Variation in the retention counts of new confirmed point sources (left) and false positives (right) with h under different SNR filtering ranges for Atlas images in dense stellar fields. The images are from Sky Regions 7 and 8. The horizontal axis represents h, while the vertical axis shows the normalized count of detected sources (relative to the maximum count). Curves in different colors correspond to different upper SNR limits. The subplot titles indicate the maximum number of detected sources.
\label{fig:4}}
\end{figure*}

\begin{figure*}[ht!]
\plotone{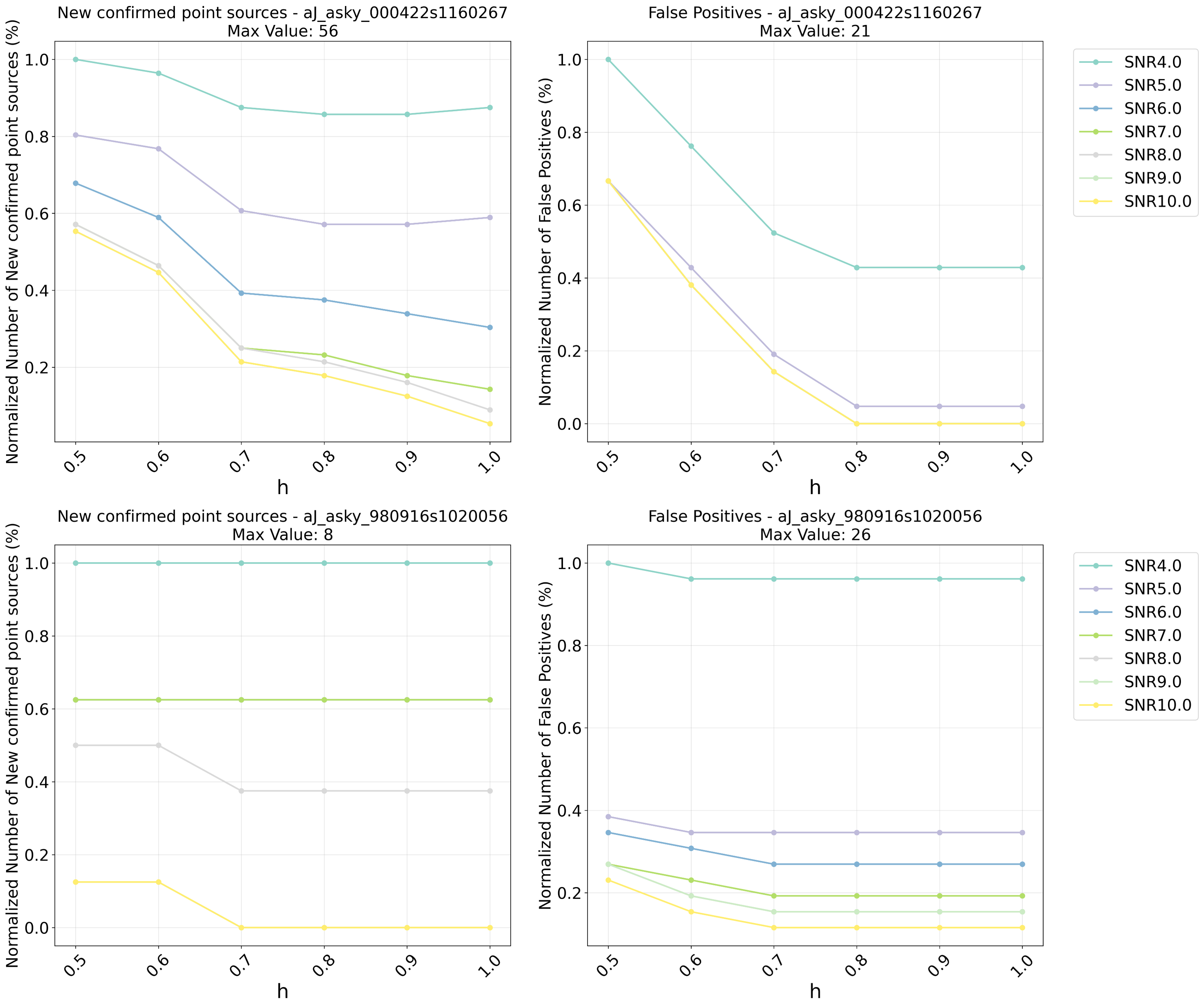}
\caption{Variation in the retention counts of new confirmed point sources (left) and false positives (right) with h under different SNR filtering ranges for Atlas images in sparse stellar fields. The images are from Sky Regions 1 and 3. Other specifications follow those of Figure \ref{fig:4}.
\label{fig:5}}
\end{figure*}

In this step, sharpness filtering is applied exclusively to low SNR sources to reduce the false positive rate. Due to variations in detection background, Atlas images with different stellar densities require analysis within distinct low SNR ranges to determine an appropriate sharpness threshold for peak sharpness screening. Detections with SNR greater than the upper limit of the SNR screening range are retained directly; only those within the SNR range whose sharpness falls below h are rejected (h is the preset sharpness threshold). To determine the optimal filtering parameters for different Atlas images in this step, the upper limits of the SNR screening range were set to $[4, 5, 6, 7, 8, 9, 10]$, with a fixed lower limit of 3, while the sharpness threshold h was tested at values of $[0.5, 0.6, 0.7, 0.8, 0.9, 1.0]$. Figure \ref{fig:3} shows the distribution of all detections with $3 \leq SNR \leq 10$ in different Atlas images. As illustrated, false positives among low SNR detections are predominantly concentrated in the region where $h \leq 0.5$, a trend particularly evident in sparse fields. Therefore, the minimum value of h was set to 0.5. Figures \ref{fig:4} and \ref{fig:5} present the retention counts of new confirmed point sources and false positives as a function of the sharpness threshold within different SNR screening ranges, for representative Atlas images of varying stellar densities. Table \ref{tab:2} presents the statistical results of the optimal parameter combinations for different Atlas images. These combinations were selected from all parameter sets with a false positive rate below $10\%$ that achieved the highest growth rate. Table \ref{tab:3} provides a comparative analysis of the growth rate and false positive rate before and after filtering. 

\begin{deluxetable*}{p{0.5cm}p{0.5cm}p{0.5cm}p{0.5cm}p{0.5cm}p{0.5cm}p{0.5cm}p{0.5cm}p{0.5cm}p{0.5cm}}
\tablewidth{0pt}
\tablecaption{Optimal filtering parameters and statistical results for the eight Atlas images \label{tab:2}}
\tablehead{
\colhead{ID} & \colhead{SNR} & \colhead{h} & \colhead{2MASS} & \colhead{Detection} & \colhead{New} & \colhead{Extended} & \colhead{False positives} & \colhead{GR($\%$)} & \colhead{FP($\%$)}
} 
\startdata
1 & 6 & 0.7 & 64 & 82 & 5 & 8 & 7 & 7.81 & 8.54  \\
2 & 5 & 0.7 & 116 & 160 & 31 & 1 & 15 & 26.72 & 9.38 \\
3 & 4 & 0.6 & 163 & 207 & 54 & 14 & 16 & 33.13 & 7.73 \\
4 & 4 & 0.9 & 216 & 192 & 5 & 0 & 12 & 2.31 & 6.25 \\
5 & 10 & 0.5 & 312 & 357 & 37 & 9 & 30 & 11.86 & 8.40 \\
6 & 6 & 0.8 & 317 & 436 & 67 & 19 & 39 & 21.14 & 8.94 \\
7 & 4 & 0.5 & 2813 & 3123 & 459 & 181 & 79 & 16.32 & 2.53 \\
8 & 4 & 0.5 & 2178 & 1560 & 475 & 139 & 58 & 30.45 & 2.66 \\
\enddata
\tablecomments{The ID corresponds to that in Table \ref{tab:1}. SNR is the upper limit for the signal-to-noise ratio screening, h is the sharpness threshold, 2MASS represents the number of point sources from the PSC in the detection region, Detection is the number of detected sources, New is the number of new confirmed point sources, Extended is the number of detected extended sources, and False Positives refers to the number of false positives. GR$\%$ and FP$\%$ are defined identically to those in Equations 7 and 8. All subsequent tables follow the same conventions.}
\end{deluxetable*}

Based on a comprehensive analysis of Table \ref{tab:2} and Figures \ref{fig:4},\ref{fig:5}, it is found that in dense stellar fields, as the selection criteria become more stringent, the number of new confirmed point sources declines much more rapidly than that of false positives. Therefore, the optimal filtering parameters for dense stellar fields are determined to be$\mathrm{SNR} \in [3,4]$ and $h = 0.5$. In regions where the total number of detected objects is less than 1000, however, the optimal filtering parameters lack a consistent value. It is noteworthy that in sparse star fields, the number of false positives declines more rapidly than that of new confirmed point sources. Moreover, when $\mathrm{SNR} \in [3,6]$ and $h = 0.7$, the decline in the numbers of both types of sources gradually slows and eventually plateaus. Accordingly, screening strategies were adjusted based on star field density: for dense star fields (object counts $>$ 2000), the focus is on maximizing the detection of new confirmed point sources; for sparse star fields (object counts $<$ 1000), the priority is to eliminate false positives as thoroughly as possible; and for regions with intermediate density, a compromise is made regarding the constraint parameters. The specific screening procedure is as follows:

1) When the number of detected objects (N) exceeds 2000, apply sharpness screening to sources within the SNR range of 3 to 4, and filter out those with a sharpness value below $h = 0.5$.

2) When N is between 1000 and 2000, apply sharpness screening to sources within the SNR range of 3 to 5, and filter out those with a sharpness value below $h = 0.6$.

3) When N is below 1000, apply sharpness screening to sources within the SNR range of 3 to 6, and filter out those with a sharpness value below $h = 0.7$.

Since the sizes of the detection regions are essentially consistent across different Atlas images in this study, the number of detected objects can be used as a proxy for the regional stellar density.

\begin{deluxetable*}{rllll}
\tablewidth{0pt}
\tablecaption{Comparison between the pre- and post-screening growth and false positive rates for the eight Atlas images. \label{tab:3}}
\tablehead{
\colhead{ID} & \colhead{pre-$GR(\%)$} & \colhead{post-$GR(\%)$} & \colhead{pre-$FP(\%)$} & \colhead{post-$FP(\%)$}
}
\startdata
1 & 21.88 & 7.81 & 63.31 & 8.54 \\
2 & 50.86 & 26.72 & 52.80 & 8.54 \\
3 & 41.72 & 33.13 & 18.73 & 9.38 \\
4 & 3.24 & 2.31 & 12.66 & 7.73 \\
5 & 33.01 & 11.86 & 22.76 & 6.25 \\
6 & 55.21 & 21.14 & 29.60 & 8.40 \\
7 & 19.27 & 16.32 & 2.48 & 2.53 \\
8 & 42.24 & 30.45 & 3.32 & 2.66 \\
\enddata
\end{deluxetable*}

Table \ref{tab:3} presents the numerical changes in the growth rate and false positive rate for different images before and after sharpness screening. It can be observed that in most image regions, this screening method demonstrates considerable effectiveness in enhancing the detection of faint point sources while reducing the false positive rate.

\begin{figure*}[ht!]
\plotone{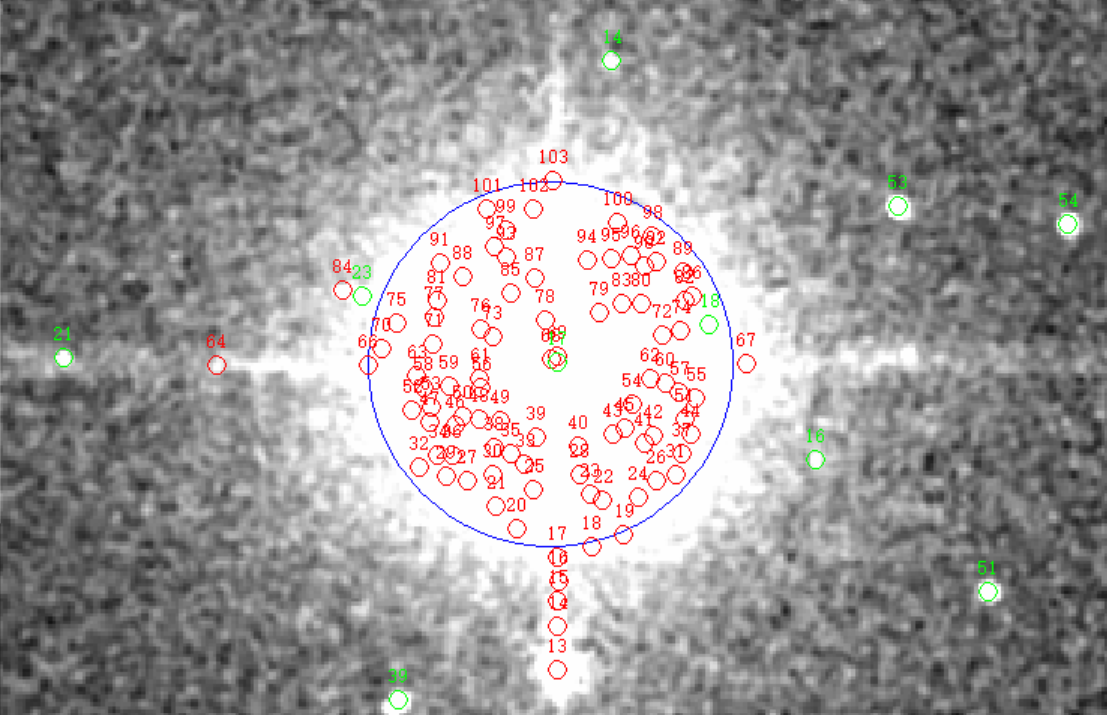}
\caption{Schematic diagram of false positives distribution around saturated sources. Green points represent 2MASS point sources, red points indicate detected false positives, and the large blue circle denotes the region within a $<57^{\prime\prime}$ radius centered on the saturated source.
\label{fig:6}}
\end{figure*}

\subsection{Processing of saturated regions} 

Local saturation is a common issue in 2MASS Atlas images and can lead to numerous spurious detections surrounding region of saturation, as shown in Figure \ref{fig:6}. Therefore, prior to compiling the final catalog of detected objects, we process such cases as follows: all detected objects within a radius of r (see Equation 9, in arcseconds) centered on the celestial coordinates of the saturated source are removed. This empirical formula, derived from observing the affected areas around multiple saturated sources, effectively eliminates most false positives induced by saturation (pixel value $>2^{13}$).

\begin{equation}
r = -2247 + 251 \cdot \ln v, \quad v > 2^{13}
\end{equation}

where r denotes the radius of the exclusion region, and $v$ represents the central pixel peak value.

\subsection{Complete pipeline} 

\begin{figure*}[ht!]
\plotone{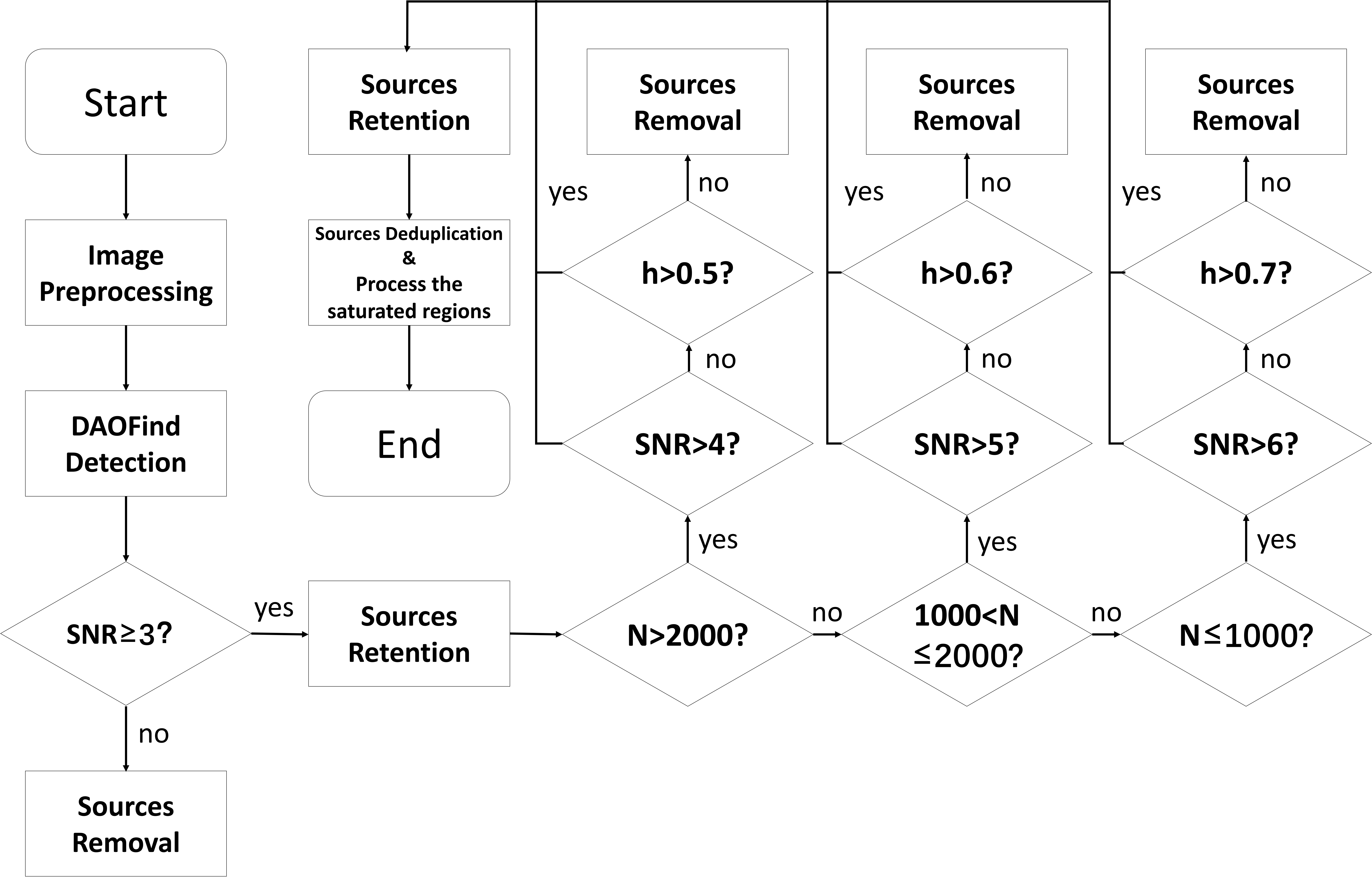}
\caption{Pipeline flowchart.
\label{fig:7}}
\end{figure*}

Figure \ref{fig:7} illustrates the complete detection pipeline. Following image preprocessing, Atlas images are processed using the unified DAOFind parameters specified in $\S2.2.1$, retaining all detections with $SNR \geq 3$. Subsequently, central sharpness screening is applied following the methodology presented in $\S2.2.2$, which utilizes distinct low SNR ranges and sharpness thresholds adapted to different stellar density regimes to reduce false positives. Finally, a simplified processing step is applied to false positives in saturated regions using the method described in $\S2.3$, while simultaneous nearest-neighbor matching with a $1^{\prime\prime}$ radius is performed to remove duplicate sources in overlapping areas of adjacent Atlas images.

\section{RESULTS AND ANALYSIS} 
In this section, the image data from the eight sky regions listed in Table \ref{tab:1} are used as a test set to obtain detection results through the complete pipeline. The corresponding statistical results are provided in Table \ref{tab:4}. As shown in Table \ref{tab:4}, the detection method for Atlas images presented in this study proves effective for most regions—it achieves a significant growth rate in point sources while maintaining the false positive rate at or below $10\%$, and also detects a considerable number of extended sources. However, as evidenced by Tables \ref{tab:3} and \ref{tab:4}, the pipeline performance in Sky Region 4 is suboptimal, exhibiting a false positive rate substantially higher than its growth rate. We attribute these detection results primarily to contamination by bright interstellar medium in this particular area. Since the pipeline demonstrates suboptimal performance for source detection in Sky Region 4, we have excluded this region's data from subsequent analyses.

\begin{deluxetable*}{p{0.5cm}p{0.5cm}p{0.5cm}p{0.5cm}p{0.5cm}p{0.5cm}p{0.5cm}p{0.5cm}}
\tablewidth{0pt}
\tablecaption{Summary of detections from the eight sky regions \label{tab:4}}
\tablehead{
\colhead{ID} & \colhead{2MASS} & \colhead{Detection} & \colhead{New} & \colhead{Extended} & \colhead{False positives} & \colhead{GR($\%$)} & \colhead{FP($\%$)}
} 
\startdata
1 & 2,080 & 2,887 & 250 & 318 & 318 & 12.02 & 10.88 \\
2 & 3,826 & 5,368 & 789 & 256 & 634 & 20.62 & 11.81 \\
3 & 7,718 & 10,701 & 1,857 & 829 & 946 & 24.06 & 8.84 \\
4 & 10,502 & 10,235 & 596 & 164 & 1,280 & 5.68 & 12.51 \\
5 & 15,323 & 18,481 & 2,440 & 481 & 1,176 & 15.92 & 6.36 \\
6 & 12,699 & 16,886 & 2,686 & 494 & 1,487 & 21.15 & 8.81 \\
7 & 15,631 & 20,939 & 4,062 & 1,102 & 1,001 & 25.99 & 4.78 \\
8 & 66,836 & 98,228 & 25,265 & 5,523 & 3,244 & 37.81 & 3.30 \\
\enddata
\end{deluxetable*}

\begin{figure*}[ht!]
\plotone{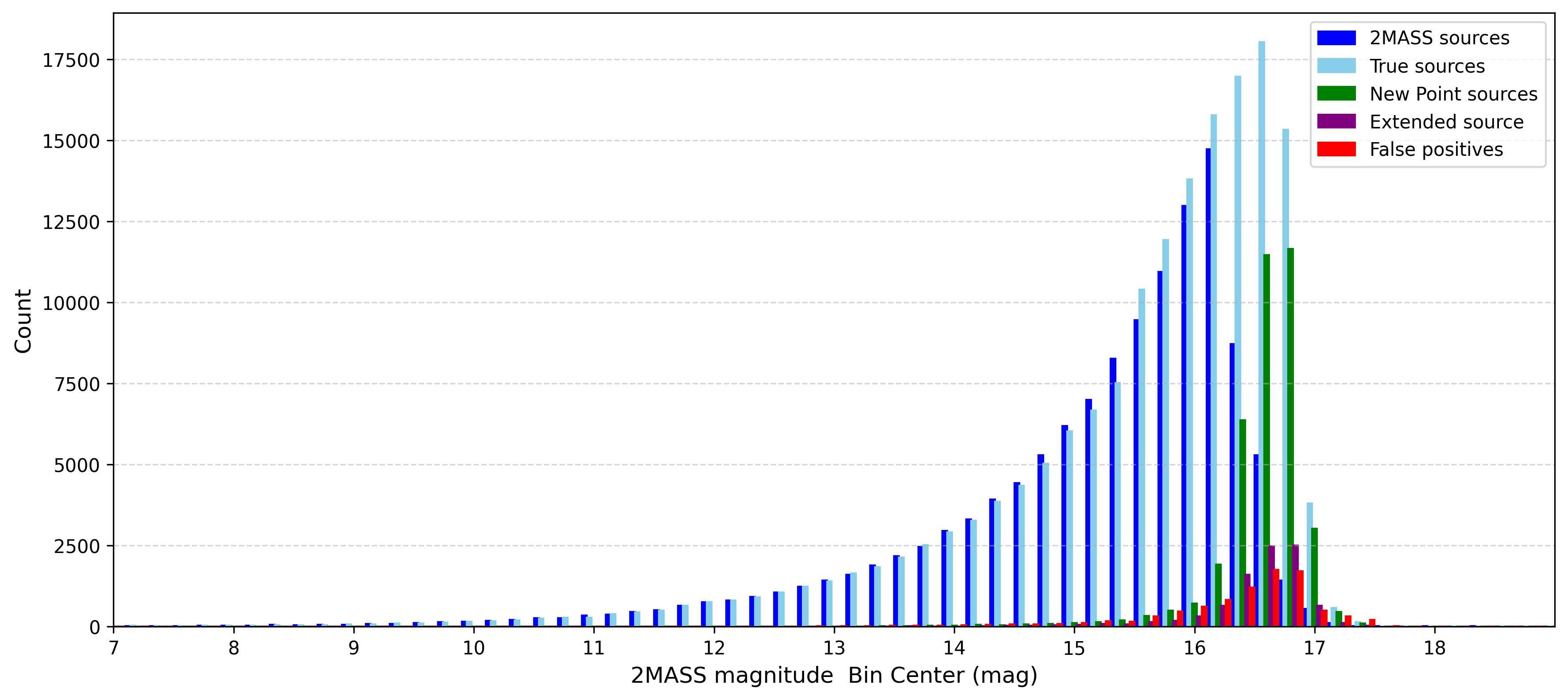}
\caption{Distribution of different source categories across 2MASS magnitude bins in the test set. Each grouped bar chart comprises five distinct color-coded categories: point sources from the PSC (dark blue); all confirmed sources (light blue); new confirmed point sources (green); extended sources (purple); and false positives (red). All magnitude bins have a fixed width of 0.2 magnitudes. The horizontal axis represents the median values of the 2MASS magnitude bins, while the vertical axis indicates the number of corresponding detected sources in each bin.
\label{fig:8}}
\end{figure*}

\begin{figure*}[ht!]
\plotone{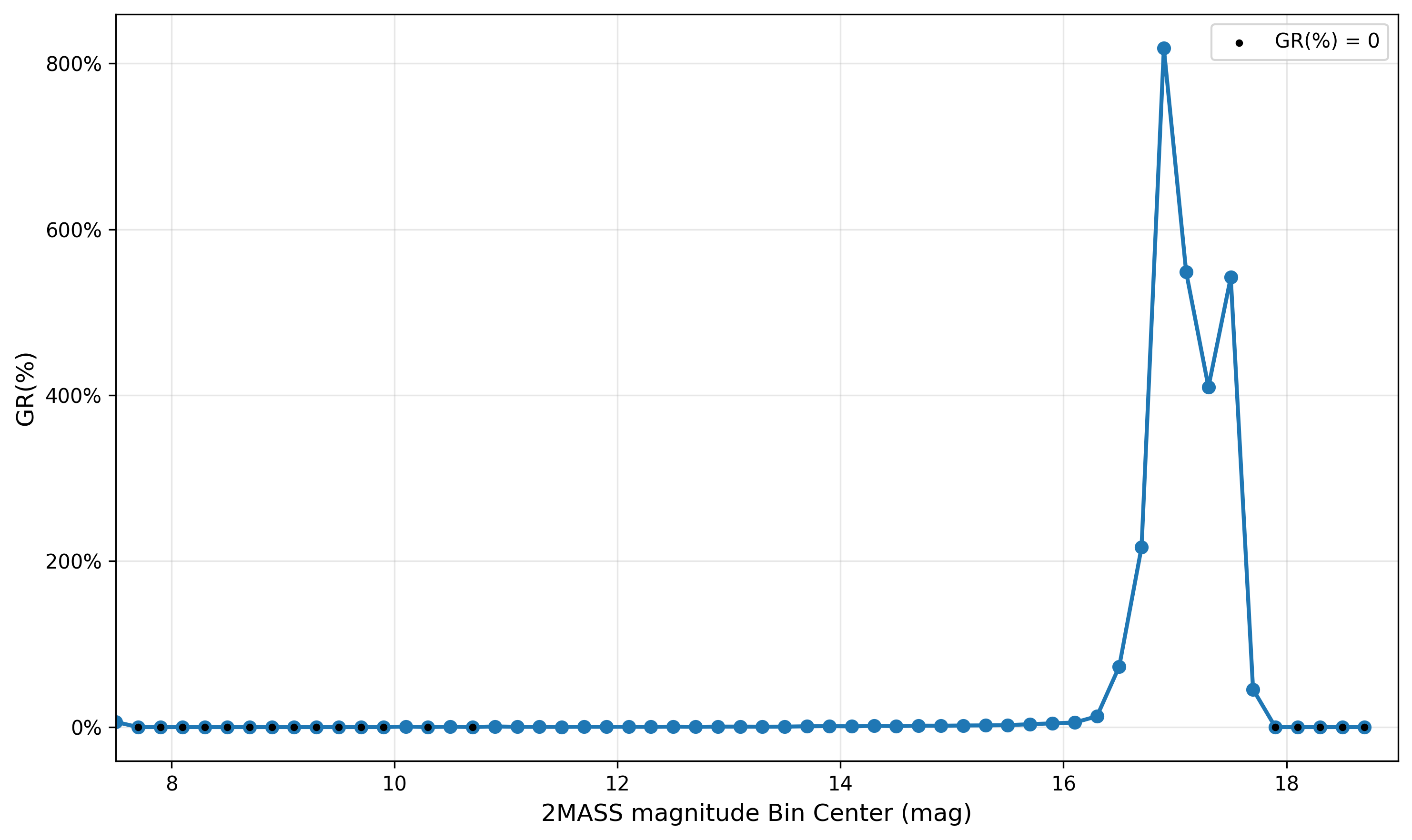}
\caption{Variation of point source growth rate across 2MASS magnitude bins in the test set. The horizontal axis follows the same convention as in Figure \ref{fig:8}, and the vertical axis shows the growth rate. The black dot marks the position where the growth rate $GR(\%)$ equals zero.
\label{fig:9}}
\end{figure*}

\begin{figure*}[ht!]
\plotone{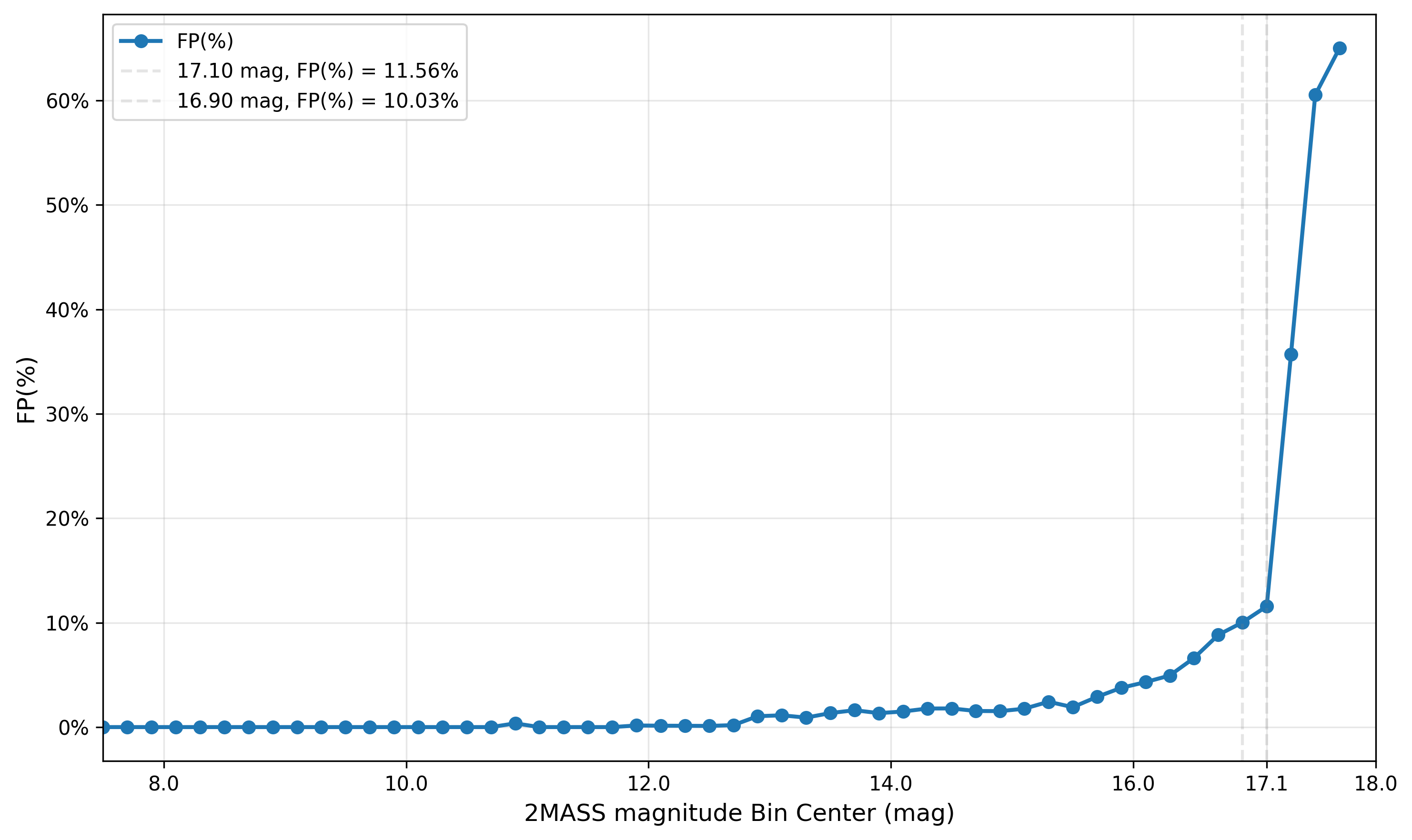}
\caption{Variation of false positive rate across 2MASS magnitude bins in the test set. The horizontal axis uses the same magnitude bins as defined in Figure \ref{fig:8}, and the vertical axis represents the false positive rate.
\label{fig:10}}
\end{figure*}

Next, we analyze the influence of magnitude on the detection results. First, the aperture magnitudes calculated from the Atlas images are converted to the 2MASS magnitude system (see Appendix A). Second, we plot the relationships between magnitude and the number of various detected sources, growth rate, and false positive rate (Figures \ref{fig:8}-\ref{fig:10}), thereby enabling a more comprehensive and detailed analysis.

From Figure \ref{fig:8}, the following preliminary conclusions can be drawn for the test set: 1) The number of the PSC J-band point sources peaks at approximately 16.20 mag. Through re-detection of Atlas images, this study has extended the peak in the number of confirmed point sources to approximately 16.60 mag. 2) New confirmed point sources, confirmed extended sources, and false positives are primarily concentrated within the $[16.40, 17.20]$ mag range. Within this interval, the quantity of new confirmed point sources significantly exceeds that of both false positives and 2MASS point sources, while a considerable number of faint extended sources are also detectable. 3) Across all magnitude bins, there exist original 2MASS point sources that remain undetected, termed “false negatives" (the phenomenon also noted in \citep{masias2013quantitative}). 4) For sources fainter than 17.2 mag, false positives constitute the majority of the detection results.

This study is dedicated to detecting faint point sources at the limit of the PSC. The growth rate and the false positive rate are the key metrics under investigation, which will be analyzed in detail as functions of the 2MASS magnitude.

Analysis of Figure \ref{fig:9} indicates an unprecedented surge in the growth rate between 16.00 and 18.00 mag, culminating in a peak exceeding $800\%$ around 16.90 mag. Outside this range, the growth rate hovers around zero. However, it is crucial to note that despite the bimodal peaks in the curve, Figure \ref{fig:8} demonstrates a precipitous drop in the number of new confirmed point sources beyond 17.20 mag—precisely between the two peaks.

Figure \ref{fig:10} shows that the false positive rate grows as the magnitudes become fainter. It remains low ($<10\%$) across the $[7.80, 16.9]$ mag range. Since this study focuses on detecting faint new confirmed point sources, the combined evidence from Figures \ref{fig:8}-\ref{fig:10} reveals that detections fainter than 17.20 mag become substantially less reliable, marked by a precipitous drop in confirmed sources and a sharp rise in false positives.

The following quantitative results for the overall growth rates ($\mathrm{GR}_{\mathrm{e}}$) and false positive rates ($\mathrm{FP}_{\mathrm{e}}$) are obtained across specified magnitude ranges: 1) Brighter than 17.20 mag, $\mathrm{GR}_{\mathrm{e}}=21.36\%$, $\mathrm{FP}_{\mathrm{e}}= 4.80\%$. 2) Entire dataset: $\mathrm{GR}_{\mathrm{e}}=21.53\%$ and $\mathrm{FP}_{\mathrm{e}}= 5.07\%$. However, for detected sources fainter than 17.20 mag, their authenticity is difficult to guarantee.

Next, we will compare the detection results from the aforementioned sky regions with the combined data from the PSC and PSRT to discuss the differences between the two source datasets in the low SNR regime. Since the PSRT contains sources of low reliability, it cannot be used as a basis for determining whether the detected sources are real astrophysical sources; therefore, we employ the VHS data as an external reference. Subsequently, according to the catalog type and reliability, the PSC and PSRT data were organized into three distinct datasets: the PSC data alone, the combined data from the PSC and PSRT, and the combined data from the PSC and the PSRT after excluding sources with reliability flags(“rel”) “E” and “F”. Here, rel is the reliability flag provided in the PSRT, where “E” indicates a reliability P $\in [20\%, 50\%]$, and “F” indicates a reliability lower than $20\%$ \citep{cutri2003explanatory}.

The brief processing steps are as follows: 1) Since duplicate sources exist either within the PSRT data itself or between the PSRT and PSC data in some sky regions, deduplication is performed on the combined PSC and PSRT dataset using a nearest-neighbor radius of $0.5^{\prime\prime}$. 2) The deduplicated 2MASS data and the pipeline-detected data from this study are separately cross-matched with the VHS data using a matching radius of $2^{\prime\prime}$. Sources in these datasets that have no counterpart in VHS are considered potential false positives. The numbers of such potential false positives in the different datasets are given in Table \ref{tab:5} below. 3) The pipeline-detected data from this study are then matched against the deduplicated 2MASS data with a matching radius of $1^{\prime\prime}$. Through steps 2 and 3, we can identify the new confirmed point sources in our detection data relative to the different 2MASS datasets. Table \ref{tab:6} presents the corresponding numbers of new confirmed point sources for each dataset. The relevant results are as follows:

\begin{deluxetable*}{llll}
\tablewidth{0pt}
\tablecaption{Cross-matching results of pipeline data and the three-class 2MASS datasets with the VHS dataset. \label{tab:5}}
\tablehead{
\colhead{Dataset} & \colhead{Sources} & \colhead{VHS-matched sources} & \colhead{VHS-unmatched sources}
} 
\startdata
PSC & 124,113 & 121,966 & 2,147 \\
PSC and PSRT & 678,573 & 290,213 & 388,360 \\
PSC and PSRT (Excluding “E” and “F”) & 197,511 & 188,190 & 9,321 \\
Pipeline Detection & 173,490 & 164,688 & 8,802 \\
\enddata
\tablecomments{“Sources" refers to the total number of sources in the dataset used. “VHS-matched sources" refers to the number of sources from the dataset that were successfully matched to a VHS source. “VHS-unmatched sources" refers to the number of sources from the dataset that were not matched to any VHS source. (Sources = VHS-matched sources + VHS-unmatched sources).}
\end{deluxetable*}

\begin{deluxetable*}{llll}
\tablewidth{0pt}
\tablecaption{Number of new confirmed point sources identified by the pipeline data relative to the three-class 2MASS datasets. \label{tab:6}}
\tablehead{
\colhead{2MASS dataset} & \colhead{PSC} & \colhead{PSC and PSRT} & \colhead{PSC and PSRT (Excluding “E” and “F”)}
} 
\startdata
Number of new confirmed point sources & 37,352 & 12,699 & 15,712
\enddata
\end{deluxetable*}

As shown in Table \ref{tab:5}, 1) a significant number of false positives exist in the PSRT data with the rel parameter marked as “E" or “F"; 2) the total number of sources detected by the pipeline is lower than that from the PSC and PSRT data after removing sources with “E" and “F" flags, while the number of false positives remains largely consistent. Furthermore, based on Table \ref{tab:6}, compared to the PSC and PSRT data after excluding “E" and “F" sources, the pipeline detection results provide 15,712 new confirmed point sources; even when compared to the full PSC and PSRT dataset, the pipeline still yields 12,699 new confirmed point sources. Thus, the results presented in this study serve as a valuable complement to the PSC and PSRT data. Additionally, in the low SNR regime, the pipeline method developed herein enhances the ability to distinguish true detections from false positives.

\section{CONCLUSIONS} 

Based on eight representative 2MASS sample regions, this study establishes optimized DAOFind parameters to detect faint point sources in Atlas images that are not included in the PSC. Following signal-to-noise ratio screening, we implement secondary filtering based on central sharpness to reduce false positives. The pipeline additionally addresses the impact of saturated regions and duplicate detections. This automated system demonstrates enhanced capability in detecting faint point sources while effectively suppressing false positives, though it shows limited applicability in sky regions affected by bright interstellar medium.

Our analysis shows that the pipeline pushes the detection limit of Atlas images from the original 16.20 mag to 16.60 mag, yielding a substantial number of new confirmed point sources. Furthermore, with 17.20 mag as the boundary, the false positive rate increases rapidly for sources fainter than this limit. For brighter sources, the false positive rate grows gradually with magnitude, peaking at no more than $11.56\%$ across different magnitude bins. The overall false positive rate remains at $4.80\%$, while achieving a growth rate of approximately $21.36\%$. More importantly, our pipeline detected point sources that were missed even by the PSRT, thereby effectively complementing both the PSC and PSRT.

Building upon the pipeline developed in this work, we will process the all-sky 2MASS Atlas images to construct an improved near-infrared source catalog. Future work will focus on: 1) performing photometric calibration for faint sources; 2) using Gaia's high precision astrometric data as a positional reference to re-calibrate the positions of near-infrared sources; 3) improving the pipeline's performance in challenging regions; and 4) ultimately producing a new, high-precision near-infrared catalog.

Based on this pipeline, subsequent work will involve processing Atlas images to create a new high-precision near-infrared star catalog. Future tasks primarily include: 1) Conducting flux calibration for faint targets; 2) Using the high-precision Gaia star catalog data as a positional reference to re-calibrate the positions of near-infrared sources; 3) Improving the pipeline's detection performance in special sky regions (e.g., Region 4 and areas near saturated sources); 4) Processing the H- and Ks-band Atlas All-Sky Release Survey images; 5) Attempting to extend the application of this pipeline to other 2MASS Atlas images and even to images from other surveys; 6) Ultimately, delivering a new high-precision near-infrared star catalog.

\begin{acknowledgments}

This work has been supported by the Strategic Priority Research Program of the Chinese Academy of Sciences (Grants No. XDA0350201, XDA0350204, and XDA0350205), the National Natural Science Foundation of China (Grants No. 12073062, 12173069, and 12473070), the Preresearch Project on Civil Aerospace Technologies funded by the China National Space Administration (Grant No. D010105), the International Partnership Program of Chinese Academy of Sciences (Grant No. 018GJHZ2023110GC), the Youth Innovation Promotion Association CAS (Certificate Number 2022259), and the Talent Plan of Shanghai Branch, Chinese Academy of Sciences (Grant No. CASSHB-QNPD-2023-016). We acknowledge the science research grants from the China Manned Space Project with Nos. CMS-CSST-2021-A12, CMS-CSST-2021-B10, and CMS-CSST-2021-B04.

This publication makes use of data products from the Two Micron All Sky Survey, which is a joint project of the University of Massachusetts and the Infrared Processing and Analysis Center/California Institute of Technology, funded by the National Aeronautics and Space Administration and the National Science Foundation. 

Based on data products created from observations collected at the European Organisation for
Astronomical Research in the Southern Hemisphere under ESO programme 179.A-2010 and made use of data from the VISTA Hemisphere survey \citep{mcmahon2013first} with data pipeline processing with the VISTA Data Flow System \citep{irwin2004vista,lewis2010pipeline,cross2012vista}

\end{acknowledgments}

%
\facilities{CTIO:2MASS, FLWO:2MASS \citep{https://doi.org/10.26131/irsa2},
ESO-VISTA:VHS \citep{https://doi.org/10.18727/archive/57}}

\software{astropy \citep{astropy:2013,astropy:2018,astropy:2022},  
          photutils \citep{larry_bradley_2025_14889440}, 
          scipy \citep{2020SciPy-NMeth}
          }


\appendix

\section{Magnitude transformation}

This step requires photometric data from the PSC. The following magnitude conversion is performed solely to facilitate integrated analysis of the detected sources in the test set, and should not be regarded as providing formally calibrated magnitudes. The equation for calculating the aperture magnitude is given by:

\begin{equation}
M_{\mathrm{cal}} = \mathrm{MAGZP}- 2.5 \times \log_{10}\mathrm{f}
\end{equation}

Here, $M_{\mathrm{cal}}$ is the aperture magnitude obtained from the Atlas images, and MAGZP is the magnitude zero point from the header of the Atlas images. In this paper, the aperture photometry radius is uniformly set to $\sigma = \mathrm{FWHM}_{\mathrm{img}}/(2\sqrt{2\ln 2})$ , and the flux 
$f$ within the aperture is calculated using the background-subtracted image data. This photometric aperture helps reduce the influence of nearby bright sources on faint detected sources. However, due to its relatively small size, it may slightly underestimate the flux of non-faint targets. To ensure consistency between the magnitudes of the detected targets and those provided in the PSC, we matched the detected targets with the 2MASS point sources. A functional fit was performed on the magnitude data of successfully matched detected point sources and their corresponding 2MASS point sources. The magnitude transformation equation is as follows:

\begin{equation}
M_{\mathrm{2MASS}} = 
\begin{cases}
-1.293 \times M_{\mathrm{cal}}^2 + 27.733 \times M_{\mathrm{cal}} - 139.634, & M_{\mathrm{cal}} \le 10.3~\mathrm{mag} \\
1.009 \times M_{\mathrm{cal}} - 1.496, & 10.3~\mathrm{mag} < M_{\mathrm{cal}} \le 12~\mathrm{mag} \\
0.980 \times M_{\mathrm{cal}} - 1.119, & 12~\mathrm{mag} < M_{\mathrm{cal}} \le 17~\mathrm{mag} \\
0.820 \times M_{\mathrm{cal}} + 1.621, & M_{\mathrm{cal}} > 17~\mathrm{mag}
\end{cases}
\end{equation}

Where $M_{\mathrm{2MASS}}$ is the fitted approximate 2MASS magnitude, and $M_{\mathrm{cal}}$ is the aperture magnitude calculated in Equation A1.


\bibliography{sample701}{}
\bibliographystyle{aasjournalv7}



\end{document}